\documentclass[conference]{IEEEtran}
\IEEEoverridecommandlockouts
\usepackage{cite}
\usepackage{amsmath,amssymb,amsfonts}
\usepackage{algorithmic}
\usepackage{graphicx}
\usepackage{textcomp}
\usepackage{xcolor}
\usepackage{tikz}
\usepackage{diagbox}
\usepackage{caption}
\usepackage{longtable}
\usepackage{multirow}
\usepackage{booktabs}
\usepackage{color}
\usepackage{bm}
\usepackage[marginal]{footmisc}

\def\BibTeX{{\rm B\kern-.05em{\sc i\kern-.025em b}\kern-.08em
    T\kern-.1667em\lower.7ex\hbox{E}\kern-.125emX}}
\begin{document}

\title{Delay-Aware Scheduling over mmWave/Sub-6 Dual Interfaces: A Reinforcement Learning Approach\\
}

\author{\IEEEauthorblockN{Ying Cao, Bo Sun, Danny H.K. Tsang}
}

\maketitle

\begin{abstract}
We consider a transmitter with mmWave/sub6 dual interfaces. Due to the intermittency of mmWave channel, the transmitter must schedule packets wisely across the interfaces to minimize the average delay by observing the system state. We use the well-known dynamic programming methods and Q-learning to find the optimal scheduling policy and investigate the influence of observing CSI on the optimal policy under different levels of knowledge of the environment. We find that only when the channel state transition model is not available, the instantaneous CSI can help in reducing system delay. 
\end{abstract}


\section{Introduction}
To enable emerging technologies such as augmented reality and connected automonous vehicles, the fifth generation cellular wireless network will utilize the massive spectrum in millimeter wave bands (above 10 GHz), which can potentially boost the wireless capacity for eMBB services and reduce the transmission delay for low-latency applications. However, the mmWave band is inherently unstable for providing reliable connections, and thus the most promising solution is to integrate the stability of sub-6 GHz and the high capacity of mmWave networks\cite{roadmap}. Standardization bodies and industry partners have also recently emphasized the importance of \textit{mmWave-$\mu$W integrated technology} as a cost-effective solution to achieve high capacity, low latency and reliability for emerging wireless applications.

Nevertheless, the intermittency and fast-changing property of the mmWave channel still induce much difficulty in channel estimation and resource allocation. Since reliable mmWave transmission necessitates that both channel estimation and resource allocation keep up with the fast channel variations, it would be much easier if a certain statistical model exists in the channel variations. However, in most cases, we do not have a statistical model of the channel. To this end, reinforcement learning is deemed to be a viable tool, by utilizing observations from the past and making decisions based upon the gained knowledge. 
\vspace{-0.2cm}
\subsection{Related Work}
It is common to utilize channel knowledge to improve resource allocation in the wireless community. However, a specific probability distribution for the channel states is usually assumed. For an unknown channel model, \cite{online_2008} first provides an online implementation of the value iteration algorithm for optimal packet scheduling; however, this method only applies to slow-varying channels. 

Recently, papers on learning a fast-changing wireless channel such as mmWave have been published. Some of these formulate the problem into a multi-armed bandit (MAB) framework since the feedback of channels are usually bandit-formed. For example, \cite{link_sel} considers the optimal rate selection problem in rapidly-changing channels, where the user only has access to bandit feedback of a successful transmission over the channels. \cite{learn_ch} studies the problem of learning channel statistics to efficiently schedule transmission in wireless networks under interference constraints. Our work differs in that we model the network as an Markov Decision Process (MDP) rather than a single state MAB problem, and thus the state transition process is more complicated than those.

Our work is inspired by \cite{yao2019}, where the authors propose the system model that we build upon. Note that \cite{yao2019} added a processing server right after the head buffer. Since the delay incurred during this process only contains the delay of reading the packet from the head buffer and of writing it to the transmission buffer, and the data transfer rate of the latest variant of memory (DDR4) is around 25GB/s, which is an order of magnitude higher than the target peak download rate of 5G (i.e., 20Gb/s \cite{itu-2020}), we assume the scheduling delay, i.e., the time from making the scheduling decision to the time the packet actually arrives at the corresponding server, is negligible. Due to this observation, we don't need a processing server after the head buffer. Moreover, we assume that the scheduler can access the mmWave server occupancy status with no delay, which is a common assumption in the literature\cite{gopalan2017}\cite{jia2017}. From the aforementioned research, intrinsically we ask the following questions:
\begin{itemize}
    \item \textit{When will channel state information (CSI) help scheduling?}
    \item \textit{If the explicit channel model is not known, could the learning algorithm perform better than the queue-length-threshold policy?}
\end{itemize}
\vspace{-0.3cm}
\subsection{Our Contributions}
Through extensive simulations, we find that under the full information of channel state transition kernel, the delay-optimal policy is still a threshold policy on queue length, which means the instantaneous CSI does not help.
However, when the channel state transition kernel is inaccessible, the presence of CSI helps further improve the delay performance.

\section{System Model}
\begin{figure}[!htbp]
	\includegraphics[scale=0.6]{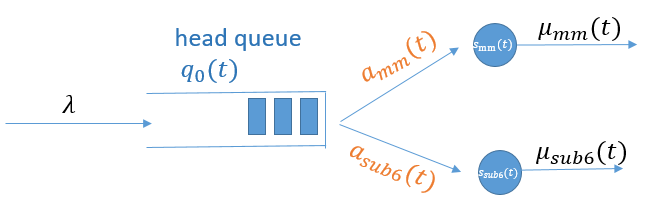}
	\caption{System model: mmWave sub-6 GHz scheduler.}
	\label{fig:model}
\end{figure}

We consider an integrated mmWave/sub-6 scheduler as shown in Fig. \ref{fig:model}. The scheduler consists of one buffer and two servers, of which one is the mmWave interface and the other is the sub-6 GHz interface. Time is divided into equal-sized slots with length $\tau$ and is indexed by $t=1,...,T$. We consider the case of non-preemptive scheduling, i.e., the packet in the server cannot be interrupted during transmission, and the scheduler has access to the mmWave CSI only when it makes the scheduling decision but no knowledge of that when the packet is being transmitted. Meanwhile, the sub-6 GHz server is rather stable in terms of service time but the service rate is much slower compared to the average service rate at the mmWave interface. Confronted with two servers, one with high average service rate but highly dynamic nature and the other one with low but stable service rate, the objective of the scheduler is to wisely make the scheduling decision for the packet(s) at the beginning of the head queue at each time slot so as to minimize the average delay.

\vspace{-0.3cm}
\subsection{Two-Layer Markov mmWave Channel Model}

\begin{figure}[!htb]
    \centering
    \includegraphics[scale=0.25]{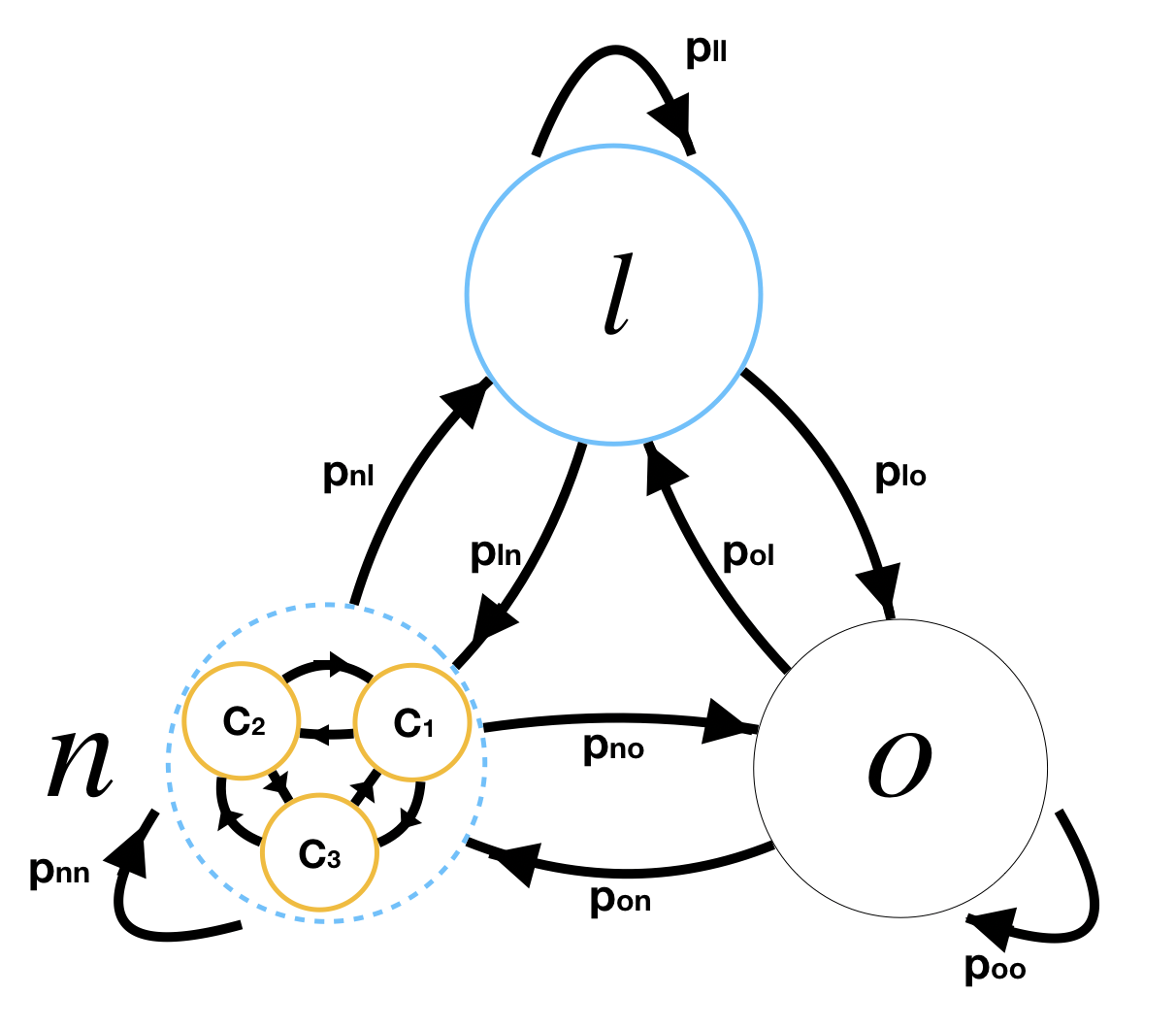}
    \caption{Two-layer mmWave channel model}
    \label{fig:ch_model}
\end{figure}

It is well-known that the mmWave link is highly dynamic and easily blocked. The more widely-used mmWave channel model in the mmWave networking community is the 3-state Markov chain with line of sight (LoS), non-line of sight (NLoS) and outage states\cite{ch_model_origin}. A more accurate finite state Markov chain (FSMC) model with 2 layers of variations is proposed in \cite{ch_model}. Here, we summarize the general workings of the model and show the diagram in Fig. \ref{fig:ch_model}.
\vspace{-0.3cm}
\subsubsection{Long-Term Link State Model}
The random process describing the transitions between macro-scale shadowing states is modeled as a Markov chain with states $st=\{l,n,o\}$, denoting LoS, NLoS and outage, respectively. We let $st(t)$ denote the link state at time $t$. The state transition kernel $\{p_{ij}\}_{i,j\in\{l,n,o\}}$ and the steady state probability $\mathbb{P}(st)$ defines the long-term link state model completely, where $p_{ij}=\mathbb{P}(st(t+1)=j|st(t)=i)$.

\subsubsection{Small-Scale Capacity Model}
The model characterizes the small-scale fading effect when the long-term link state is fixed. The capacity is calculated as 
\begin{align}
	C = W\log(\frac{P_{TX}G_{mm}}{N_0 W}),
\end{align}
where $G_{mm}$ is the squared magnitude of the mmWave channel gain. Here we do not consider the channel matrix but only focus on the magnitude of the channel gain, since in many cases, for example, admission control, the SNR is sufficient for making decision\cite{jmc}.

The small-scale channel capacity model is another Markov chain within state $C$. For NLoS state, the channel capacity is quantized into $N$ levels, and the channel capacity in the $i$th level is denoted by $c^{n}_i$, $i\in\{1,...,N\}$. For LoS state, the channel capacity is quantized into $L$ levels and the channel capacity in $i$th level is $c^l_i$, $i\in \{1,...,L\}$. For outage state, the channel capacity $c^o = 0$. Let $C(t)$ denote the link capacity at time $t$. For the NLoS state, the small-scale FSMC is defined by the state transition model $\{q^n_{ij}\}_{i,j\in \{1,...,N\}}$ and $\mathbb{P}^{(n)}(C)$, where $q^n_{ij} = \mathbb{P}(C(t+1)=c^n_j|C(t)=c^n_i)$ and $\mathbb{P}^{(n)}(c^n_j)$ is the steady state probability of having capacity $c^n_j$ in link state $n$. Likewise, for the LoS state, the subscript changes from $n$ to $l$. For the sake of brevity, Fig. \ref{fig:ch_model} shows the case for $N=3$, $L=1$ and omits the self-transition loop for the small-scale capacity model due to the space limit. 

Denote the combined channel state as $ch(t)=(st(t),C(t))$. The general two-layer state transition model can be derived as 
$$
\mathbb{P}(ch(t+1)=(j,c^j_k)|ch(t)=(i,c^i_m))=
\left\{
\begin{tabular}{ll}
     $p_{ii}q^i_{mk}, j=i$ \\
     $p_{ij}\mathbb{P}^{(j)}(c^j_k), j\neq i$.
\end{tabular}
\right.
$$ 
Compared with the two-layer mmWave channel, the sub-6 GHz channel is much more stable, and thus we assume that it is a single-state channel with simple small-scale capacity level dynamics.
\vspace{-0.2cm}
\subsection{System Dynamics}
\subsubsection{MDP Formulation} \label{MDP}
Let $A(t)$ denote the random new packet arrivals at the end of the $t$-th scheduling slot, which is assumed to be i.i.d. Poisson distributed over scheduling slots with mean $\mathbb{E}[A]=\lambda$. $X(t)$ denotes the packet size of the packet in the front of the head queue at the beginning of the $t$-th time slot and is assumed to be i.i.d. exponential distributed over scheduling slots with mean $\Bar{X}$. The departure rate is state-dependent.
\begin{itemize}
	\item State Space: $\mathbf{s}(t)=(\bm{q}(t), ch_{mm}(t))\in \mathcal{S}$, where $\bm{q}(t)=(q_0(t), s_{mm}(t), s_{sub6}(t))$ and $ch_{mm}(t)=(st_{mm}(t),C_{mm}(t))$. Note that we introduce the subscript for mmWave channel in order to differentiate it from the sub6-6 GHz channel. Let $q_0(t)\in \{0,1,...\}$ denote the head queue length and $s_i(t)\in \{0,1\},i\in \{mm,sub6\}$ denote the occupancy of the mmWave server and the sub-6 GHz server, respectively. Given the state $\mathbf{s}(t)$, the average departure rate at the mmWave interface is given by $\mu_{mm}(\bm{s}(t))=\mathbb{E}[\frac{R(\bm{s}(t))}{X(t)}|\bm{s}(t)]=\frac{R(\bm{s}(t))}{\Bar{X}}$ and the probability of a packet departure there at the $t$-th slot can be approximated by $\mu_{mm}(\bm{s}(t))\tau$ $\footnote{To show this, we need another assumption: $\mu_{mm}(\bm{s}(t))\tau\ll1$, the detailed proof is given in \cite{lau2009}.}$.
	\item Action Space: $\bm{a}(t)=(a_{mm}(t), a_{sub6}(t))$. The action space is state-dependent:
	\begin{itemize}
	    \item $\bm{q}=(0,s_{mm},s_{sub6})$:
	    $\mathcal{A}=\{(0,0)\}$
	    \item $\bm{q}=(1,s_{mm},s_{sub6})$:\\ $\mathcal{A}=\{(a_{mm},a_{sub6})|a_{mm}+s_{mm}<2, a_{sub6}+s_{sub6}<2, a_{mm}+a_{sub6}<2\}$
	    \item $\bm{q}=(q_0\geq 2,s_{mm},s_{sub6})$:\\ $\mathcal{A}=\{(a_{mm},a_{sub6})|a_{mm}+s_{mm}<2, a_{sub6}+s_{sub6}<2\}$
	\end{itemize}
	A decision rule, $\pi:\mathcal{S} \rightarrow \mathcal{A}(\bm{s})$, is a function mapping from the state space $\mathcal{S}$ to the action space $\mathcal{A}(\bm{s})$.
	\item Transition Kernel: $\mathbb{P}(\mathbf{s}(t+1)|\mathbf{s}(t),\mathbf{a}(t))$. The generic expression of this kernel is: 
	\[
	\left \{
		\begin{tabular}{ll}
			$\lambda \tau \mathbb{P}(ch_{mm}(t+1)|ch_{mm}(t))$,  $C_1$ \\
			$\mu_{mm} \tau \mathbb{P}(ch_{mm}(t+1)|ch_{mm}(t))$, 
			$C_2$ \\
			$\mu_{sub6} \tau \mathbb{P}(ch_{mm}(t+1)|ch_{mm}(t))$, $C_3$ \\
			$1-(\lambda+\mu_{mm}+\mu_{sub6})\tau\mathbb{P}(ch_{mm}(t+1)|ch_{mm}(t))$,
			$C_4$,
		\end{tabular}
	\right.
	\]
	where $\mu_{mm}=\mu_{mm}(\bm{s}(t))$ and
	\begin{align*}
	    &C_1=
    	\left\{
        	\begin{aligned}
        	&q_0(t+1)=q_0(t)-a_{mm}(t)-a_{sub6}(t)+1\\ &s_1(t+1)=\min(s_1(t)+a_{mm}(t),1)\\ &s_2(t+1)=\min(s_2(t)+a_{sub6}(t),1),\\
        	\end{aligned}
    	\right. \\
    	&C_2=
    	\left\{
        	\begin{aligned}
        	&q_0(t+1)=q_0(t)-a_{mm}(t)-a_{sub6}(t)\\ &s_1(t+1)=\max(\min(s_1(t) +a_{mm}(t)-1,1),0)\\ 
        	&s_2(t+1)=\min(s_2(t)+a_{sub6}(t),1),\\
        	\end{aligned}
    	\right. \\
    	&C_3 =
    	\left\{
        	\begin{aligned}
        	&q_0(t+1)=q_0(t)-a_{mm}(t)-a_{sub6}(t)\\ &s_1(t+1)=\min(s_1(t)+a_{mm}(t),1)\\ 
        	&s_2(t+1)=\max(\min(s_2(t)+a_{sub6}(t)-1,1),0),\\
        	\end{aligned}
    	\right. \\
    	&C_4 =
    	\left\{
        	\begin{aligned}
        	&q_0(t+1)=q_0(t)-a_{mm}(t)-a_{sub6}(t)\\ &s_1(t+1)=\min(s_1(t)+a_{mm}(t),1)\\ 
        	&s_2(t+1)=\min(s_2(t)+a_{sub6}(t)-1,1).\\
        	\end{aligned}
    	\right.
	\end{align*}
	Note that $C_1$, $C_2$ and $C_3$ are the queue dynamics for different events, which correspond to the arrival, the departure from the mmWave server and the departure from the sub6 GHz server, respectively. $C_4$ means nothing happens. \\
\end{itemize}
\vspace{-0.3cm}
\subsubsection{Two levels of CSI knowledge}
The queue state information (QSI) can be accessed in real time, which is a common assumption in the literature\cite{ying2010}\cite{javidi2007}. Regarding the CSI, although it is known that the complete instantaneous mmWave CSI $ch_{mm}(t)$ is usually unavailable, knowledge of the channel transition kernel and the large-scale CSI $st_{mm}(t)$ can be acquired easily in some cases, where dynamic programming methods can be used to obtain the optimal policy. For example, when the environment is rather static, the channel transition matrix can be estimated offline; Advanced channel tracking technologies \cite{ch_est} can be utilized to estimate $st_{mm}(t)$. In other cases, when we do not know the channel transition matrix in advance, traditional dynamic programming methods will fail and learning methods will surge. Based on this divergence, we divide the possible channel state information into 2 levels.
\begin{itemize}
	\item Known $\mathbb{P}\{ch_{mm}(t+1)|ch_{mm}(t)\}$, known $ch_{mm}(t)$: Full information of the stationary channel state transition model is known. Additionally, we assume the scheduler can observe the complete instantaneous CSI, in order to investigate how CSI can help scheduling.
	\item Unknown $\mathbb{P}\{ch_{mm}(t+1)|ch_{mm}(t)\}$, known $st_{mm}(t)$: The scheduler has no information about the channel state transition kernel except the instantaneous large-scale CSI.
\end{itemize}
\vspace{-0.2cm}
\subsection{Problem Formulation}
We are concerned with minimizing the average delay incurred in the system. Based on Little's law, the average delay experienced by one packet is proportional to the average number of packets in the system, thus the problem is
\begin{equation}
	\begin{aligned}
	\nonumber &\underset{\mathbf{\pi}}{\min}
	&&\underset{T \rightarrow \infty}{\limsup}
	\frac{1}{T}\mathbb{E}^\mathbf{\pi} \sum_{t=1}^{T}q_t,
	\end{aligned}
\end{equation}
where $q_t=q_0(t)+s_{mm}(t)+s_{sub6}(t)$.
\section{Techniques}
\subsection{Dynamic Programming}
For the known channel transition matrix case, the optimal policy can be calculated by exact dynamic programming methods. Both relative value iteration (RVI) and linear programming (LP) are implemented in this work. The key update step in RVI for infinite horizon average cost problem is:
$$
h_{i+1}(\bm{s}) = (Th_{i})(\bm{s})-(Th_{i})(\bm{s}_0),
$$
where $(Th_{i})(\bm{s})=\underset{\bm{a}\in \mathcal{A}(\bm{s})}{\min} \left[ C(\bm{s})+\sum_{\bm{s}'\in \mathcal{S}} P(\bm{s}'|\bm{s},\bm{a})h_{i}(\bm{s}'))\right]$ and $\bm{s}_0$ is a fixed state. $C(\bm{s})$ is the cost function, which is set to $q_0+s_{mm}+s_{sub6}$ in our problem. As $i\rightarrow \infty$, $h_{i}(\bm{s})$ will converge to the differential cost $h^*(\bm{s})$ for state $\bm{s}$ w.r.t. the optimal average cost for all states $(Th^*)(\bm{s}_0)$. The optimal policy $\pi(\bm{s})=\underset{\bm{a}\in \mathcal{A}(\bm{s})}{\text{argmin }} \left[ C(\bm{s})+\sum_{\bm{s}'\in \mathcal{S}} P(\bm{s}'|\bm{s},\bm{a})h_{i}(\bm{s}'))\right]$.\\ For LP, we solve the optimization problem as follows:
\begin{equation}
	\begin{aligned}
	\nonumber &\underset{\bm{q}(\bm{s},\bm{a})}{\min}
	&& \sum_{\bm{s}\in \mathcal{S}}\sum_{\bm{a}\in \mathcal{A}(\bm{s})} q(\bm{s},\bm{a})C(\bm{s}) \\
	&s.t. && \sum_{\bm{a}\in\mathcal{A}(\bm{s})} q(\bm{s},\bm{a})=\sum_{\bm{s}'\in \mathcal{S}}\sum_{\bm{a}\in\mathcal{A}(\bm{s}')} q(\bm{s}',\bm{a})P(\bm{s}|\bm{s}',\bm{a}), \forall \bm{s} \\
	&&& \sum_{\bm{a}\in\mathcal{A}(\bm{s})} q(\bm{s},\bm{a})=1, \forall \bm{s} \\
	&&& q(\bm{s},\bm{a})\geq 0, \forall \bm{s}, \bm{a},
	\end{aligned}
\end{equation}
where $q^*(\bm{s},\bm{a})$ denotes the state-action probability under the optimal policy. It is direct to see that the policy output by linear programming can be randomized due to the probability distribution, while value iteration is guaranteed to output a deterministic optimal policy since the optimal value function is unique.
\vspace{-0.3cm}
\subsection{Q-learning}
Q-learning is a classic model-free reinforcement learning (RL) method proposed originally in \cite{q_learn}. The essence is to approximate the optimal action value function using learned action value function from sampled rewards:
$$
\begin{aligned}
    Q_{t+1}(\bm{s}_t,\bm{a}_t)&=(1-\alpha)Q_t(\bm{s}_t,\bm{a}_t)\\
    &+\alpha(R(\bm{s}_t)+\gamma \underset{\bm{a}\in \mathcal{A}(\bm{s}_{t+1})}{\max} Q_t(\bm{s}_{t+1},\bm{a})),
\end{aligned}
$$
where $\alpha \in (0,1)$ is the learning rate and $\gamma \in(0,1)$ is the discounted ratio. It can be proven that when $\gamma\rightarrow 1$, the system converges to the average reward case. $R(\bm{s})$ is the reward observed in state $\bm{s}$, which is set to $\frac{1}{q_0+s_{mm}+s_{sub6}}$. In this work, since we have an explicit expression on the queueing dynamics, the queue size will not increase significantly as long as the system satisfies the queue stability condition, the state space can be handled only by tabular Q-learning. However, note that if considering a wireless network, the state space may be too large for tabular methods, and thus one can seek the help of function approximation. This is beyond the scope of this paper.
 
\section{Simulation: Settings and Results}
Since the techniques we use are different under the two information levels, we divide the simulation section into two parts. 
\subsection{Known $\mathbb{P}(ch_{mm}(t+1)|ch_{mm}(t))$, known $ch_{mm}(t))$} \label{known_tran_mod}\label{result_1}
Under this channel information level, we implemented the value iteration and linear programming methods to find the optimal policy. The case without considering channel state information is studied in \cite{yao2019}. Since our objective is to investigate the effect of CSI on the scheduling policy, for comparison, we use a similar system model as in \cite{yao2019}, which adds a buffer before the mmWave server. Nevertheless, it is further verified by the output optimal policy that, if the instantaneous CSI is known and there is no scheduling delay, the buffer before the mmWave server is not needed.

\begin{figure}[!h]
	\includegraphics[scale=0.5]{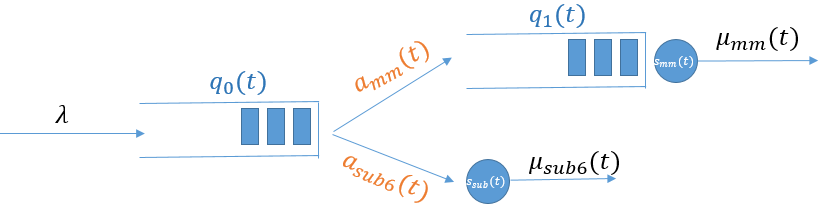}
	\caption{System model: modified mmWave-sub 6GHz scheduler.}
	\label{fig:model_2}
\end{figure}
\vspace{-0.3cm}
\subsubsection{Simulation Parameters}
\begin{itemize}
    \item State Space: $q_0\in \{0,1,...,6\}$, $q_1\in \{0,1,...,5\}$, $s_{mm},s_{sub6}\in \{0,1\}$, $st_{mm} \in \{l,n,o\}$. We follow the parameters shown in Fig. \ref{fig:ch_model} and $C_{mm}\in \{c^l,c^n_1,c^n_2,c^n_3,c^o\}$. Here we normalize the maximum channel capacity to 1 and set $(c^l,c^n_1,c^n_2,c^n_3,c^o)=(1,0.05,0.004,0.002,0)$. The state space size is 840. We simulate the mmWave channel state transition kernel based on the sub-6 GHz channel. Assume the sub-6 GHz channel has two states denoted by $st_{sub6} \in \{\text{bad},\text{good}\}$. The detailed parameters of the mmWave/sub6 CSI model are as follows:
    \begin{itemize}
    		\item sub-6: $\mathbb{P}(st_{sub6}=\text{bad})$ = 0.2, $\mathbb{P}(st_{sub6}=\text{good})$ = 0.8.
    		\item mmWave: see Table \ref{tab:my_label}.
    	\end{itemize}
    \begin{minipage}[b]{.50\textwidth}
        \centering
        \vspace{0.3cm}
        \captionof{table}{Conditional distribution of $C_{mm}$}	\label{tab:my_label}
    	\begin{tabular}{|c|c|c|}
			\hline	\diagbox{$C_{mm}$}{$\mathbb{P}(C_{mm}| st_{sub6})$}{$st_{sub6}$} & bad & good \\\hline
			$c^l$ & 0.1 & 0.7 \\ \hline
			$c^n_1$ & 0.15 & 0.15 \\ \hline
			$c^n_2$ & 0.15 & 0.05 \\ \hline
			$c^n_3$ & 0.15 & 0.05 \\ \hline
			$c^o$ & 0.45 & 0.05 \\ \hline
		\end{tabular}
		\vspace{0.3cm}
    \end{minipage}
	\item Action Space: $a_{mm} \in \{0,1\}, a_{sub6} \in \{0,1\}$, and the action space size is 4.
	\item As shown in Table \ref{table: ch_gain_pro}, the small-scale mmWave channel gain follows Gaussian processes with different means and variances in different channel states.\\
\begin{minipage}[b]{.50\textwidth}
	\centering
	\vspace{0.3cm}
	\captionof{table}{Small-scale mmWave channel gain model}
	\label{table: ch_gain_pro}
	\begin{tabular}{|c|c|c|} 
		\hline
		LoS & NLoS($i$th level) & outage \\
		\hline
		$N(c^l,0.001)$ & $N(c^n_i,0.1)$ & 0\\
		\hline
	\end{tabular}
	\vspace{0.3cm}
\end{minipage}
    \item The arrival rate $\lambda$ is normalized to 1 pkt/s, and the departure rate is shown in Tables  \ref{table:2} and \ref{table:1}. We can see that under undesirable mmWave channel conditions, the departure rate is slower than the sub-6 GHz interface given the same power. The mean packet size $\Bar{X}$ is set to 500 kbits. In order to avoid introducing irrelevant variables, the transmission power at both interfaces, denoted by $P_{mm}$ and $P_{sub6}$ are set to 30 W.
\begin{minipage}[b]{.50\textwidth}
	\centering
	\vspace{0.3cm}
	\captionof{table}{Departure rate for sub-6 GHz interface}
	\label{table:2}
	\begin{tabular}{|c|c|} 
		\hline
		$st_{sub6}$ & $\mu_{sub6}(pkts/s)$\\
		\hline
		bad & 0.99 \\
		good & 1.45 \\ [1ex] 
		\hline
	\end{tabular}
\end{minipage}
\begin{minipage}[b]{.50\textwidth}
	\centering
	\vspace{0.3cm}
	\captionof{table}{Departure rate for mmWave interface}
	\label{table:1}
	\begin{tabular}{|c|c|} 
		\hline
		$C_{mm}$ & $\mu_{mm}(pkts/s)$ \\
		\hline
		$c^l$ & 49.54 \\
		$c^n_1$ & 13.22 \\
		$c^n_2$ & 1.64 \\
		$c^n_3$ & 0.84 \\
		$c^o$ & 0 \\
		\hline
	\end{tabular}
	\vspace{0.3cm}
\end{minipage}
Although the values are not from real data, the construction is based on verified facts:
\begin{enumerate}
    \item When the mmWave is in the LoS state, the correlation between mmWave and sub-6 GHz is strong\cite{out_of_band}.
	\item The mmWave channel in the LoS state is dozens of times the data rate of sub-6 GHz.
\end{enumerate}
\end{itemize}
\subsubsection{Simulation Results}
\begin{enumerate}
	\item State Space Reduction: Using LP, we can calculate the occurrence probability for all state-action pairs, as shown in Fig.  \ref{fig:occur_prob}. After projecting it to the state space, we can find the recurrent states, which are the states of the state-action pairs with non-zero probability.
\begin{figure}[h!]
	\centering
	\includegraphics[scale=0.5]{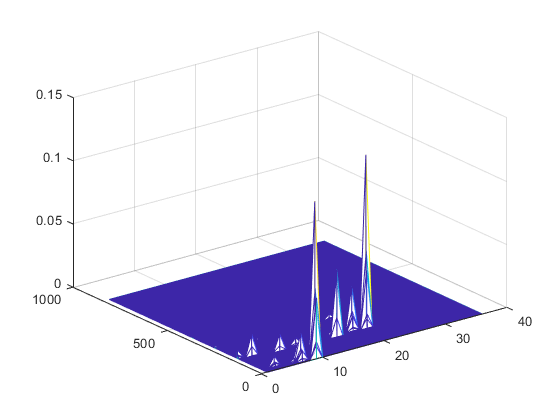}
	\caption{Occurrence probability of state-action pairs}
	\label{fig:occur_prob}
\end{figure}

After carefully examining the recurrent states, it is found that the states with $q_1>1$ never appear, which confirms that the buffer before the mmWave server is not needed. To be specific, in terms of queue state $(q_0,q_1,s)$, the recurrent states are $(n,0,0),(n,0,1),(n,1,0)$, $n\in\{0,1,...,5\}$. The reason why states with $q_0=6$ do not appear is that the arrival rate is limited. To show this, we increase the buffer size, and the recurrent states do not change. 
\begin{table}[h!]
	\centering
	\caption{Optimal policy}
	\label{tab:opt_policy}
	\begin{tabular}{|c|c|c|}
		\hline
		Queue State & CSI & Action\\ 
		($q_0,s_{mm},s_{sub6}$) & ($C_{mm},st_{sub6}$) & ($a_{mm},a_{sub6}$) \\
		\hline
		(0,x,x) & (x,x) & (0,0) \\ \hline
		\multirow{6}{*}{(1,0,0)} & ($c^o$,x) & (0,1) \\ \cline{2-3}
		~ & ($c^n_3$,bad) & (1,0) \\ \cline{2-3}
		~ & ($c^n_3$,good) & (0,1) \\ \cline{2-3}
		~ & ($c^n_2$,x) & \multirow{3}{*}{(1,0)} \\ \cline{2-2}
		~ & ($c^n_1$,x) & ~ \\ \cline{2-2}
		~ & ($c^l$,x) & ~ \\ \hline
		\multirow{5}{*}{
			\shortstack{($n$,0,0) \\ $2 \leq n \leq 6$}} & ($c^o$,x) & (0,1) \\ \cline{2-3}
		~ & ($c^n_3$,x) & \multirow{2}{*}{(1,1)} \\ \cline{2-2}
		~ & ($c^n_2$,x) & ~ \\ \cline{2-3}
		~ & ($c^n_1$,x) & \multirow{2}{*}{(1,0)} \\ \cline{2-2}
		~ & ($c^l$,x) & ~ \\ \hline
		\multirow{5}{*}{
			\shortstack{($n$,0,0) \\ $ n \geq 7$}} & ($c^o$,x) & (0,1) \\ \cline{2-3}
		~ & ($c^n_3$,x) & \multirow{4}{*}{(1,1)} \\ \cline{2-2}
		~ & ($c^n_2$,x) & ~ \\ \cline{2-2}
		~ & ($c^n_1$,x) & ~ \\ \cline{2-2}
		~ & ($c^l$,x) & ~ \\ \hline
		\multirow{5}{*}{\shortstack{($n$,1,0) \\ $1\leq n \leq 5$}} & ($c^o$,x) & \multirow{3}{*}{(0,1)} \\ \cline{2-2} 
		~ & ($c^n_3$,x) & ~ \\ \cline{2-2}
		~ & ($c^n_2$,x) & ~ \\ \cline{2-3}
		~ & ($c^n_1$,x) & \multirow{2}{*}{(0,0)} \\ \cline{2-2}
		~ & ($c^l$,x) & ~ \\ \hline
		\multirow{5}{*}{\shortstack{($n$,1,0) \\ $ n \geq 6$}} & ($c^o$,x) & \multirow{5}{*}{(0,1)} \\ \cline{2-2} 
		~ & ($c^n_3$,x) & ~ \\ \cline{2-2}
		~ & ($c^n_2$,x) & ~ \\ \cline{2-2}
		~ & ($c^n_1$,x) & ~ \\ \cline{2-2}
		~ & ($c^l$,x) & ~ \\ \hline
		
		\multirow{2}{*}{\shortstack{($n$,0,1) \\ $n \geq 1$}} & ($c^o$,x) & (0,0) \\ \cline{2-3}
		~ & (x,x) & (1,0) \\ \hline
		($n$,1,1),$n\geq 1$ & (x,x) & (0,0)\\ \hline
	\end{tabular}
\end{table}
\item Optimal Policy Structure: 

Table \ref{tab:opt_policy} shows the optimal policy. Note that when different states correspond to the same optimal action, we use "x" to represent all the possible and unspecified state components for that action. From Table \ref{tab:opt_policy}, we can see that the scheduling decision depends on the departure rate of each interface. When $(q_0,q_1,s)=(1,0,0)$, the policy schedules packets to the mmWave interface when the mmWave channel offers a higher rate than sub-6 GHz. When $(q_0,q_1,s)=(n,0,0), n \geq 2$, if the departure rates at the mmWave and sub-6 GHz are comparable, the policy schedules packets to both interfaces, but when the mmWave can offer a much higher departure rate, sub-6 GHz is put aside.

When $(q_0,q_1,s)=(n,1,0), n \geq 1$, the policy operates in a similar way as previously: it will send packets to sub-6 GHz only when the rate of the mmWave interface degenerates to the same level as that of sub-6 GHz.

Interestingly, we find that even if we have knowledge of the instantaneous CSI, the optimal policy is still the threshold type w.r.t. the queue length. In other words, the policy is not improved by the observed CSI.
\end{enumerate}

\subsection{Unknown $\mathbb{P}(ch_{mm}(t+1)|ch_{mm}(t))$, known $st_{mm}(t)$}
Since we do not have the exact channel transition model in this case, the value iteration and linear programming methods cannot apply. It is known that model-free methods in RL are suitable for tackling problems without knowledge of the model. In this part, we implemented Q-learning to learn the unknown channel transition kernel under the system model in \ref{fig:model} and compare the performance with the queue-length-threshold policy.
\subsubsection{Simulation Parameters}
\begin{itemize}
	\item State Space: $q_0 \in \{0,1,...,10\}$, $s_{mm},s_{sub6} \in \{0,1\}$, $st_{mm}\in \{l,n,o\}$.
	\item Action Space: $a_{mm}=\{0,1\}$, $a_{sub6}=\{0,1\}$.
	\item To simulate the mmWave channel, we use the transition model in \cite{mdp_model}, which characterizes an urban scenario where the dominant link is NLoS:
	$$\mathbb{P}(st_{mm}(t+1)|st_{mm}(t)) = 
	\begin{bmatrix}
		0.55 & 0.3 & 0.15 \\
		0.01 & 0.8 & 0.19 \\
		0.38 & 0.40 & 0.22
	\end{bmatrix}$$
	The average sojourn time in seconds for each state obeys $[t_{l}:t_{n}:t_{o}]=[5:25:3]$. The arrival rate $\lambda=60$ pkts/s. The mean packet size $\Bar{X}$ is set to 500 kbits. The parameters for the small-scale channel model are the same as \ref{known_tran_mod}.
\end{itemize}

\subsubsection{Simulation Results}
Under different channel models, the average time needed to transmit a packet at the mmWave interface is in Table \ref{table: mmwave_rate2}.
\begin{minipage}[c]{.5\textwidth}
	\centering
	\vspace{0.3cm}
	\captionof{table}{Average packet departure time at mmWave}
	\label{table: mmwave_rate2}
	\begin{tabular}{|c|c|c|c|c|}
		\hline
		$st_{mm}$ & $l$ & $n$ & $o$ & Average\\
		\hline
		$t(ms)$ & 3.2 & 20.4 & 32 & 19.79 \\
		\hline
	\end{tabular}
	\vspace{0.2cm}
\end{minipage}

\begin{figure}
    \centering
    \includegraphics[scale=0.6]{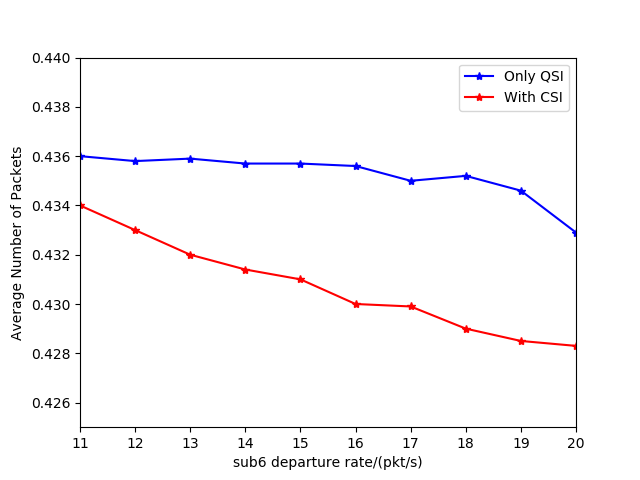}
    \caption{Performance with different sub-6 GHz rate}
    \label{fig: sub-6 rate}
\end{figure}

In Fig. \ref{fig: sub-6 rate}, we compare the performance of the queue-length-threshold policy that uses queue state information (QSI) only and the converged policy output by Q-learning that uses both CSI and QSI.
The reason that the sub-6 GHz departure rate is set in $[11,20]$ pkts/s is to ensure that the system is still in the stability region. It could be seen that the inclusion of CSI indeed helps reduce the average delay, which confirms our intuition. Compared with the results in \ref{result_1}, where the optimal policy is still queue length based, the impact of instantaneous CSI information on the scheduling policy is different. The reason could be that, the channel statistical model is higher-order information compared to the instantaneous CSI. However, since this high-order information is usually not available in real life, we show that under this case, it is of benefit to use instantaneous CSI.
Due to the slow convergence of the Q-learning method, however, the algorithm may only apply to the static environment, where the channel state transition model is stationary. Thus, a possible extension of our work is to consider a more adaptive algorithm to learn the environment faster.

\section{Conclusion}
In the forthcoming 5G era, there are emerging applications with more stringent delay and reliability requirements, and mmWave-$\mu$W integrated technology is considered as a promising solution. This paper considers the dual-interface scheduling problem in a mmWave/sub6 integrated transmitter and investigates the role of different levels of CSI in the performance of the policy. It is found that the instantaneous CSI is only helpful to the policy when the statistical knowledege of the channel is not available, where the inclusion of CSI indeed further reduces the average delay based on the delay-optimal policy with only QSI. Hopefully, the findings can help the system designers decide when it is necessary to take the instantaneous CSI into account for resource allocation.

\begin{align}
    \beta_j=
\end{align}
\end{document}